\documentclass[aip,amsmath,amssymb,reprint]{revtex4-1}
\usepackage{tikz,hyperref}
\usetikzlibrary{arrows,shapes}

\draft 

\DeclareMathOperator{\eps}{\varepsilon}
\begin{document}
\title{A new class of chimeras in locally coupled oscillators with small-amplitude, high-frequency asynchrony and large-amplitude, low-frequency synchrony}

\author{Tasso J. Kaper}
\email[]{tasso@math.bu.edu}
\affiliation{Department of Mathematics and Statistics, Boston University, Boston, MA 02215, USA}

\author{Theodore Vo}
\email[]{theodore.vo@monash.edu}
\affiliation{School of Mathematics, Monash University, Clayton, Victoria 3800, Australia}

\date{\today}
\begin{abstract}	\label{sec:abstract}
Chimeras are surprising yet important states in which domains 
of decoherent (asynchronous) and coherent (synchronous) oscillations
co-exist. 
In this article,
we report on the discovery 
of a new class of chimeras, 
called {\it mixed-amplitude chimera states},
in which the structures, amplitudes, and frequencies 
of the oscillations differ substantially
in the decoherent and coherent regions.
These mixed-amplitude chimeras
exhibit domains of decoherent small-amplitude
oscillations (phase waves) 
coexisting with domains of stable and coherent
large-amplitude 
or mixed-mode oscillations.
They are observed in a prototypical bistable partial differential equation
with spatially homogeneous kinetics 
and purely local, isotropic diffusion.
New bifurcations are identified in which the mixed-amplitude chimeras 
emerge from, or are annihilated in, common large-amplitude solutions. 
Also, key singularities, folded nodes and folded saddles, 
arising commonly in multi-scale, bistable systems
play important roles,
and these have not previously been studied in systems with chimeras. 
The discovery of these mixed-amplitude chimeras
is an important advance for understanding some
processes in neuroscience, pattern formation, and physics which involve
both small-amplitude and large-amplitude oscillations.
It may also be of use for understanding some aspects
of EEG recordings 
from animals that exhibit unihemispheric slow-wave sleep.
\end{abstract}


\maketitle

\begin{quotation}
A striking phenomenon in pattern formation occurs when groups of identical
oscillators exhibit domains of decoherent oscillations coexisting with
complementary domains of coherent dynamics. These states are known as
chimera states, named after animals from Greek mythology, which have body
parts from different known animals. However, unlike the mythological
beasts, chimera states arise ubiquitously in many areas of physics,
chemistry, neuroscience, and engineering. This article reports on the
discovery of a completely new class of chimeras. The structures,
amplitudes, and frequencies of the oscillations differ substantially in
the decoherent and coherent domains. 
Underlying these chimeras are folded singularities, which are well-known to be generating mechanisms for complex oscillatory dynamics. 
The discovery of this new class of chimeras was made on a
proto-typical bistable partial differential equation. These new chimeras
have the potential to make useful applications, including to 
unihemispheric slow wave sleep, during which the sleeping cerebral hemisphere
exhibits synchronous high-amplitude, low-frequency rhythms characteristic
of sleep, while the other hemisphere displays the decoherent
low-amplitude, high-frequency rhythms characteristic of wakefulness.
\end{quotation}

\section{Introduction \label{sec:Intro}}

Discovered in rings of identical phase oscillators \cite{KB2002,AS2004}, 
chimera states are fascinating patterns 
which exhibit co-existing domains of decoherent (asynchronous) 
and coherent (synchronous) oscillations. 
They occur broadly in models and experiments
in physics, neuroscience, chemistry, and biology,
including in systems of coupled oscillators,
integro-differential equations, and partial differential equations
\cite{OA2008,BPR2010,L2010,L2015,
OMHS2011,
TNS2012,HMRHOS2012,MTFH2013,NTS2013,SZHK2014,
YPR2014,ZKS2014,OPHSH2015,PA2015, CCFGR2016,
KHSKK2016,LD2016,NRM2017,O2018,TRTSE2018,KMMG2018,ABE2019,
SA2021,Zhang2021,SS2014,KHK2018,OMT2008}.
Coupling of the identical systems may be non-local, global, local, 
or via delay or noise.
Reviews and classifications are given in
Refs.~\onlinecite{PA2015,KHSKK2016,O2018}.

In this article, 
we report on a new class of chimeras 
--{\it mixed-amplitude chimeras}-- 
which exhibit regions of decoherent small-amplitude
oscillations (SAOs) and complementary regions of coherent large-amplitude
relaxation oscillations (LAOs) or coherent mixed-mode oscillations (MMOs).
(The latter are combinations of LAOs and SAOs
\cite{K1995,RW2007,RW2007,BKR2008,DGKKOW2012,HKWS2011}.) 
The decoherent and coherent dynamics differ qualitatively and quantitatively,
with amplitudes separated by at least one order of magnitude and
frequencies widely separated. 
The SAOs are phase waves of high frequency, 
whereas the LAOs and MMOs jump between distinct
states and are of low frequency. 

The existence of mixed-amplitude chimeras is surprising, since the LAOs
and MMOs are robust states in bistable systems. 
Also, as shown here, in parameter
regimes adjacent to the chimera regime large-amplitude states invade the
SAO regions, pushing out the decoherent SAOs there. Hence, the discovery
that decoherent SAOs can coexist with coherent LAOs and MMOs 
is of substantial scientific interest.

Along with the minimal chimeras
observed in a pair of symmetrically-coupled
Langyel-Epstein oscillators
\cite{ABE2019},
these mixed-amplitude chimeras
are the first observations of chimeras
in which the structures, amplitudes,
and frequencies of the oscillations
differ by orders of magnitude
in the decoherent and coherent regions.

In addition, the new chimeras 
will be important for applications. 
Many bistable systems 
robustly exhibit SAOs, LAOs, and MMOs.
See for example Refs.~\onlinecite{KS2009,W1997,vS2003,N2002,RG2021,AFS2004,F1960,NAY1962,HH1952,KS2009,M2006,AP1996,BKR2008,DGKKOW2012,HKWS2011,RW2007,RW2007,K1995,LRE1990,EM2008,DBSK2007,Wang2010}.
Among these,
models of various brain rhythms, cardiac rhythms, chemical waves, cell
cycle transitions, and cell signaling with positive-feedback loops
\cite{EM2008,DBSK2007,Wang2010,AFS2004,RG2021,LRE1990}, share structural
features of the model here.
In particular, it is shown here that a pair of folded singularities,
known as folded nodes and folded saddles,
which arise in systems exhibiting MMOs,
are important mechanisms responsible 
for creating the mixed-amplitude chimeras.

Also, a prominent application for which chimeras have been suggested 
as suitable models is unihemispheric slow-wave sleep (USWS)
\cite{AS2004,Motter2010,Ramlow2019,Haugland2015,Ma2010}. 
In certain aquatic mammals and birds, one cerebral hemisphere sleeps 
and the other remains awake \cite{Rattenborg2000,Rattenborg2006}. 
This is illustrated by EEG recordings from bottlenose dolphins
during USWS \cite{Rattenborg2000}, 
in which one hemisphere exhibits 
synchronous high-amplitude, low-frequency oscillations 
characteristic of sleep \cite{SMS1993}, 
whilst the other shows 
desynchronized low-amplitude, high-frequency oscillations 
characteristic of wakefulness.
We find here that the main properties of mixed-amplitude chimeras,
with co-existing regions of coherent LAOs and MMOs 
and regions of decoherent SAOs 
are similar, at least in broad strokes.
Of course, this is still far from establishing
a possible role for mixed-amplitude chimeras in USWS.
Instead, the existence of mixed-amplitude chimeras
at least demonstrates that chimeras are relevant
when the amplitudes and frequencies of the coherent 
and decoherent oscillations are widely different,
something that has not been accounted for by the existing
examples and theory of chimeras.
Moreover, the mechanisms responsible for creating the mixed-amplitude
chimeras (namely, the folded singularities of bistable systens)
can be robust to small heterogeneities.

This article is organized as follows.
In Section ~\ref{sec:fvdPol-PDE},
we study the prototypical bistable PDE 
with spatially-homogeneous reaction terms
and purely local, isotropic diffusion,
on which the mixed-amplitude chimeras were discovered.
Also, we present the main dynamical properties
and bifurcations of these chimeras.
In Section~\ref{sec:foldedsingularities},
we analyze the folded singularities
which are the mechanisms responsible
for allowing the identical oscillators
to exhibit both the decoherent SAOs
and the coherent MMOs,
{\it i.e.,}
the mechanisms responsible
for the creation of the mixed-amplitude chimeras.
In Section~\ref{sec:DHB},
the regions of SAO dynamics are examined in more detail,
and it is reported that the recently-discovered
phenomenon of delayed Hopf bifurcations
plays an important role.
In Section~\ref{sec:coh-coh},
we study the bifurcation of these chimeras
to a new type of pattern which consists
of co-existing regions of coherent MMOs and coherent SAOs.
Conclusions and discussion are presented 
in Section~\ref{sec:conclusions}.
Finally, 
Section~\ref{sec:openquestions} contains some open questions.

\section{Mixed-amplitude chimeras in a prototypical bistable PDE \label{sec:fvdPol-PDE}}
The new mixed-amplitude chimera states are presented
for a canonical multi-scale bistable PDE:
the forced van der Pol equation 
with local diffusion
\cite{KS2009,vdP1920,VKKR1987,SE1988,KG1995,FLB2002,GHW2003,SW2004,
HB2012,VW2015,BDGKKV2016},
\begin{equation}
\label{eq:fvdPol-PDE}
\begin{split}
u_t &= v - u^2 - \tfrac{1}{3} u^3 + D_u u_{xx} \\
v_t &= \eps( a - u + b \cos \theta)  + D_v v_{xx} \\
\theta_t &= \eps \omega.
\end{split}
\end{equation}
The activator (or voltage) $u$ is fast, with bistable nullcline
\cite{form-fvdPol}. The inhibitor (or recovery) $v$ is slow, and $0 < \eps
\ll 1$ measures the scale separation in the reaction terms,
as is common in bistable systems. The excitation threshold $a$,
forcing amplitude $b$, frequency $\omega$, and diffusivities $D_u$ and
$D_v$ are spatially homogeneous, so that the kinetics and diffusion are
identical and purely local at each point $x \in [0,1]$. 
The boundary conditions are zero flux.
Also, $\omega = 1$
unless stated otherwise so that the forcing period is $T = \tfrac{2\pi}{\eps}$. 
The numerical methods and typical choices of initial data
are described
in the Appendix.

\begin{figure}[h!]
\centering
\includegraphics[width=\columnwidth]{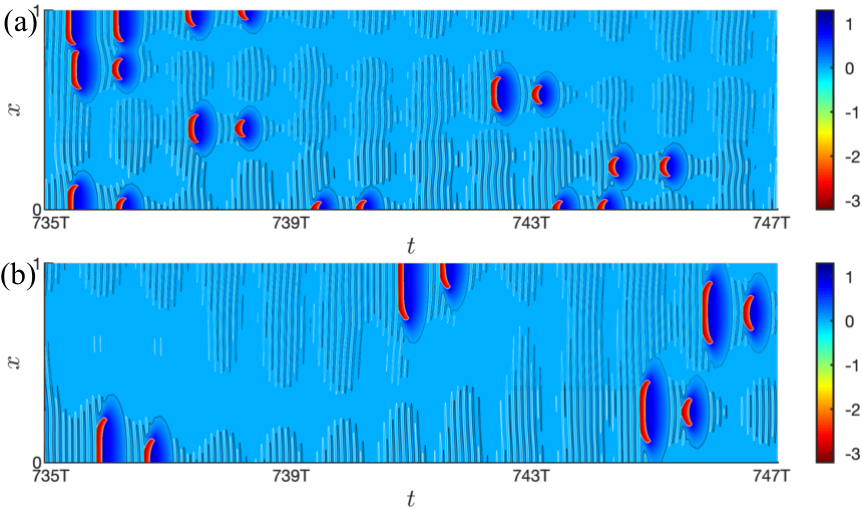}
\caption{
Mixed-amplitude chimeras of \eqref{eq:fvdPol-PDE}.
Coherent MMOs
(red regions, followed by navy blue shadows)
occur at random points
and random times.
All of the other regions
exhibit decoherent SAOs (medium blue).
Here, $a=-0.001$,
$b=0.005$,
$\eps=0.01$.
(a) $(D_u,D_v)=(0.45,4) \cdot(\Delta x)^2$,
(b) $(D_u,D_v)=(1.1,11) \cdot(\Delta x)^2$.
Contour lines at $u=\pm 2.5 \sqrt{\eps}$.
}
\label{fig:typical}
\end{figure}

\begin{figure}[h!]
\centering
\includegraphics[width=\columnwidth]{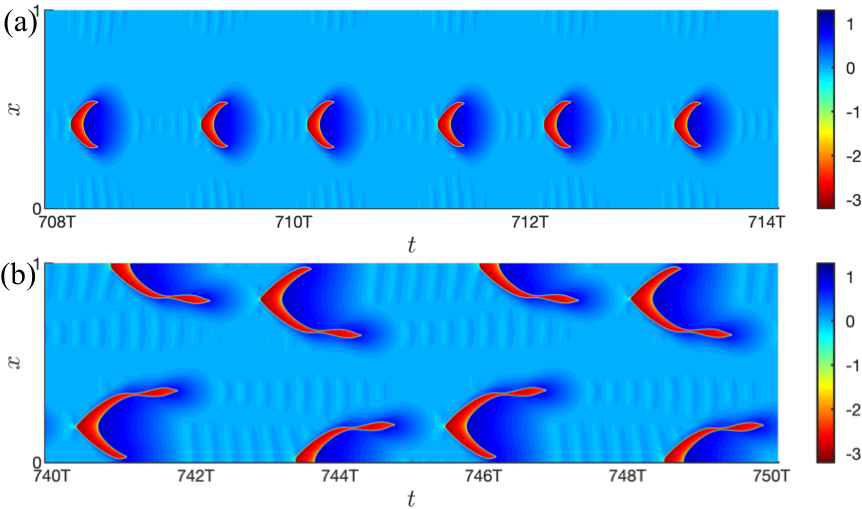}
\caption{
Mixed-amplitude
chimeras
of breathing, turbulent type.
(a) A fixed central band of coherent $2T$-periodic MMOs,
and decoherent SAOs 
in the complementary regions.
$(D_u,D_v) = (1.4,9)\cdot(\Delta x)^2$.
Kinetics parameters as in Fig.~\ref{fig:typical}.
(b) Coherent $5T$-periodic MMOs,
and decoherent SAOs centrally.
$(D_u,D_v) =
(1.7,9)\cdot(\Delta x)^2$ and $b=0.002$.}
\label{fig:representatives}
\end{figure}

The mixed-amplitude chimeras
of \eqref{eq:fvdPol-PDE}
exhibit coexisting
regions of decoherent SAOs
and coherent MMOs.
The SAOs are classical phase waves
about
the depolarized state ($u=0$).
Diffusion decoheres them. 
The MMOs
consist of coherent LAOs
between the hyperpolarized state (near $u=-3$)
and depolarized state,
alternating with coherent SAOs.
The MMOs
occur either at random points and times
(Fig.~\ref{fig:typical})
or at fixed locations and time-periodically
(Fig.~\ref{fig:representatives}),
depending on $(D_u,D_v)$.

\begin{figure}[htbp]
\centering
\includegraphics[width=\columnwidth]{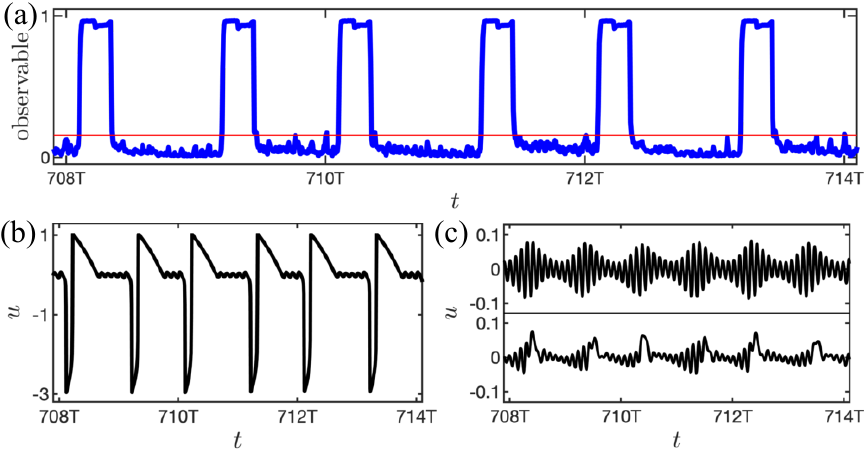}
\caption{
(a) Coherence measures $g_0(t)$ (blue)
and $h_0$ (red)
of the chimera
in Fig.~\ref{fig:representatives}(a).
Time series of $u$ in the
(b) coherent MMO region ($x=0.45$),
(c) decoherent SAO domain ($x=0.1, 0.21$).
The amplitudes
and frequencies
differ significantly.
}
\label{fig:classification}
\end{figure}

These mixed-amplitude chimeras are breathing, turbulent chimeras in the
classification scheme presented in Ref.~\onlinecite{KHSKK2016}. 
The spatial correlation 
$g_0(t) = \int_0^\delta g(|\hat{D}|,t)\, d|\hat{D}|$
was measured.
Here, $g$ is the normalized probability density of $|\hat{D}|$,
where $\hat{D} = (\Delta x)^2 D$ 
is the scaled discrete Laplacian, 
$\delta = 0.01 D_m$
is the threshold below which the profile 
is regarded as coherent,
and $D_m$ is the absolute maximal curvature.
For \eqref{eq:fvdPol-PDE},
$g_0(t)$
exhibits a pair of fast jumps 
(one up and one down) for each spatially coherent MMO.
See Fig.~\ref{fig:classification}(a).
Moreover,
these pairs of fast jumps 
occur at random times for the chimeras in Fig.~\ref{fig:typical} and
time-periodically for those in Fig.~\ref{fig:representatives} (see
Fig.~\ref{fig:classification}(a)). 
Moreover, between jumps, $g_0(t)$ fluctuates randomly,
due to the decoherent SAOs, 
remaining well inside the interval $(0,1)$, 
Hence, the solution is a breathing, turbulent chimera.

The temporal correlation 
$ h_0 = \sqrt{ \int_\gamma^1 h(|\rho|)\, d|\rho| }$
of Ref.~\onlinecite{KHSKK2016}
was also measured.
Here, $\rho_{ij}$ is the correlation coefficient 
for the time series at $x_i$ and $x_j$, 
and $h$ is the probability distribution function of $|\rho|$. 
We find that, over large times, $h_0$
converges to zero when the MMO clusters
occur randomly. 
This is consistent with the
observations that the positions of the coherent clusters change in time.
In contrast, 
for the chimeras in which the coherent MMO clusters
are fixed in space,
we find that $h_0$ is a non-zero constant 
inside $(0,1)$.
See Fig.~\ref{fig:classification}(a).

\begin{figure}[h!]
\centering
\includegraphics[width=\columnwidth]{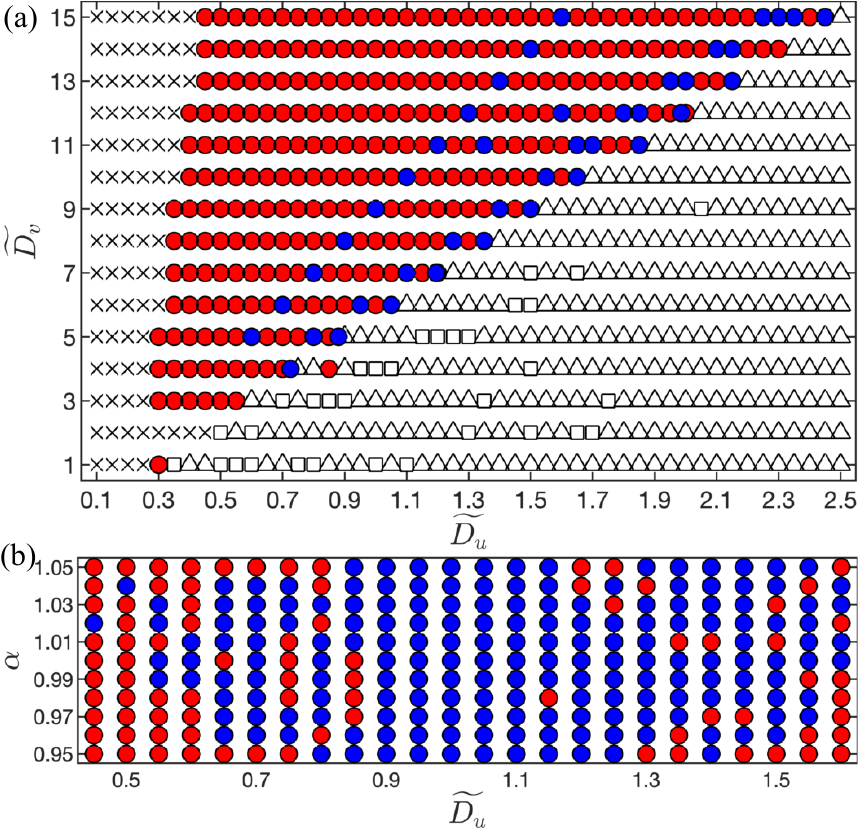}
\caption[Mixed-amplitude chimera 2-parameter plane]
{Mixed-amplitude chimeras exist for $D_u < D_v$
($D_i = {\widetilde{D_i}} (\Delta x)^2, i=u,v$).
(a) {\tikz \draw[black,fill=red] (0,0) circle (.75ex);}
Breathing turbulent chimeras with randomly-occurring coherent MMOs.
{\tikz \draw[black,fill=blue] (0,0) circle (.75ex);}
Breathing chimeras with time-periodic, spatially-fixed coherent MMOs.
$\times$ Sharp interface solutions.
$\bigtriangleup$ Spatially uniform, time-periodic MMOs.
$\Box$ Trigger waves exhibiting time-periodic annihilation and
nucleation of MMOs.
Kinetics as in Fig.~\ref{fig:typical} 
(b) In the strip ${\widetilde{D_v}} = (1 / \sqrt{\eps}) {\widetilde{D_u}} - 1
= 10 {\widetilde{D_u}} - \alpha$, for $\alpha \in [0.95,1.05]$,
many chimeras have time-periodic MMOs.
For each set of parameters,
\eqref{eq:fvdPol-PDE} was integrated
for a total time 
of $750T$.
}
\label{fig:DuDv-bifurcationdiagram}
\end{figure}

The mixed-amplitude chimeras 
of \eqref{eq:fvdPol-PDE} are observed robustly
where $D_u$ is sufficiently less than $D_v$;
see Fig.~\ref{fig:DuDv-bifurcationdiagram}. 
The typical chimeras 
(marked by red circles)
have coherent MMOs which occur at random points and at random times, 
with decoherent SAOs in all other regions,
just like the chimeras shown in Fig.~\ref{fig:typical}. 
There are also wedges 
in the $D_u-D_v$ parameter space
in which the chimeras have coherent MMOs
that are time-periodic and spatially fixed (blue circles),
examples of which are 
given by the chimeras shown in Fig.~\ref{fig:representatives}.
Moreover, in all of these chimeras,
the decoherent SAOs
occupy an increasing portion of
$[0,1]$ as $D_u$ decreases,
within this regime.

\begin{figure}[h!]
\centering
\includegraphics[width=\columnwidth]{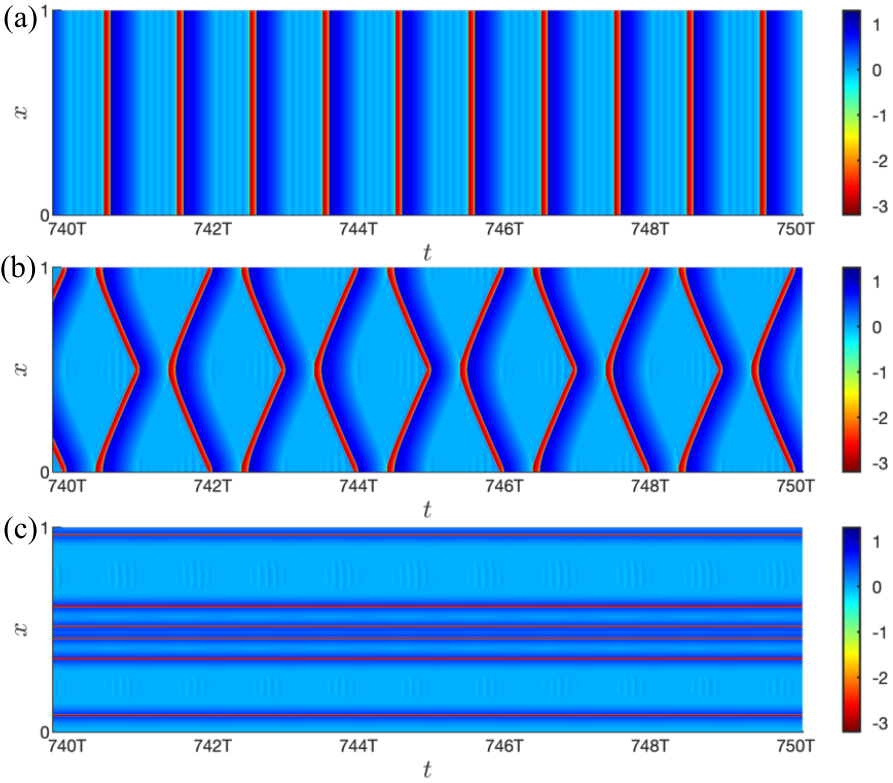}
\caption[Triangle+square]{(a) 
A representative spatially homogeneous, $T$-periodic MMO state
of \eqref{eq:fvdPol-PDE}
of the type denoted by $\bigtriangleup$
in Fig.~\ref{fig:DuDv-bifurcationdiagram}.
Here, ${\widetilde{D_u}} = 1$ and ${\widetilde{D_v}} = 5$.
(b) A representative trigger wave exhibiting time-periodic
annihilation and nucleation of MMOs,
of the type denoted by $\Box$
in Fig.~\ref{fig:DuDv-bifurcationdiagram}.
Here, ${\widetilde{D_u}} = 1.7$ and ${\widetilde{D_v}} = 7$.
(c) A representative sharp-interface solution of the type denoted by $\times$ in Fig.~\ref{fig:DuDv-bifurcationdiagram}.
Here, ${\widetilde{D_u}} = 0.15$ and ${\widetilde{D_v}} = 3$.
}
\label{fig:homogeneousMMO+TW}
\end{figure}

At the right edge 
of the mixed-amplitude chimeras regime
in Fig.~\ref{fig:DuDv-bifurcationdiagram}, 
the chimeras bifurcate 
to spatially-uniform, time-periodic MMOs (Fig.~\ref{fig:homogeneousMMO+TW}(a)) 
and to trigger waves (Fig.~\ref{fig:homogeneousMMO+TW}(b))
\cite{BDE1995}. 
Here, $D_u$ is no longer small enough relative to $D_v$.
The SAO regions are displaced (in part or entirely) by MMOs and trigger waves.
The trigger wave patterns, such as in Fig.~\ref{fig:homogeneousMMO+TW}(b), can exhibit annihilation points where pairs of traveling waves of MMOs moving in opposite directions collide and eliminate each other in the interior of the spatial domain. This annihilation event is typically followed some time later by a nucleation point at the same spatial location where a pair a trigger waves emerge and radiate away from the nucleation point in opposite directions.

Further, we observe that the presence
of the spatially coherent MMOs and trigger waves
in this parameter regime immediately adjacent
to the regime of mixed-amplitude chimeras
makes the existence of these chimeras
all the more striking.
In mixed-amplitude chimeras, 
the decoherent SAOs
resist the invasion of the stable LAO/MMOs.

At the left edge 
of the mixed-amplitude chimera regime
in Fig.~\ref{fig:DuDv-bifurcationdiagram},
the chimeras bifurcate
from sharp-interface solutions \cite{N2002}. 
As shown in Fig.~\ref{fig:homogeneousMMO+TW}(c),
these sharp interface solutions
have steep jumps at
fixed locations connecting the depolarized and hyperpolarized states.
Then, on the spatial intervals between the jumps,
these solutions exhibit time-periodic SAOs.

Overall, as shown in Fig.~\ref{fig:DuDv-bifurcationdiagram},
it is also a novelty of mixed-amplitude chimeras 
that they bifurcate 
from large-amplitude solutions,
which are ubiquitous in bistable systems, and
not from homogeneous or drift states.

Mixed-amplitude chimeras also exhibit multi-stability and
symmetry-breaking. For example, the chimera state 
in Fig.~\ref{fig:representatives}(a) 
coexists with three other stable states
of \eqref{eq:fvdPol-PDE}
for the same parameters. 
There is a co-existent spatially-homogeneous SAO state.
Also, there are two other
mixed-amplitude chimeras, 
one of which resembles the reverse image, with
coherent MMOs along the boundaries and a central strip of decoherent SAOs,
and the other of which 
has time-periodic coherent MMOs
both in the central region, 
as the chimera in Fig.~\ref{fig:representatives}(a),
and near the boundaries,
as well as decoherent SAOs everywhere else.
Furthermore, symmetry-breaking occurs
through bifurcations to chimeras 
in which the widths of the MMO regions
vary time-periodically.

\section{Folded singularities as a mechanism for mixed-amplitude chimeras
\label{sec:foldedsingularities}}

A distinct mechanism,
structurally different from that in known chimeras,
is at the heart of the mixed-amplitude chimeras.
There is a pair of commonly-occurring singularities,
known as folded nodes (FNs) and folded saddles (FSs),
in the kinetics of bistable PDEs.
These singularities
arise naturally in the 
van der Pol oscillator
\eqref{eq:fvdPol-PDE} \cite{GHW2003,VW2015,BDGKKV2016},
and in 
neuroscience, physics, chemistry, and electrical engineering models
\cite{BBE1991,RW2007,HKWS2011,TTVWB2011,DGKKOW2012,WMR2013,KAA2014,MKP2014,RRW2015,ABE2019,BBU2020,SW2004,W2005,RW2007,W2005,WMR2013,KAA2014,WMR2013,BBU2020}. 

\begin{figure}[h!]
\centering
\includegraphics[width=\columnwidth]{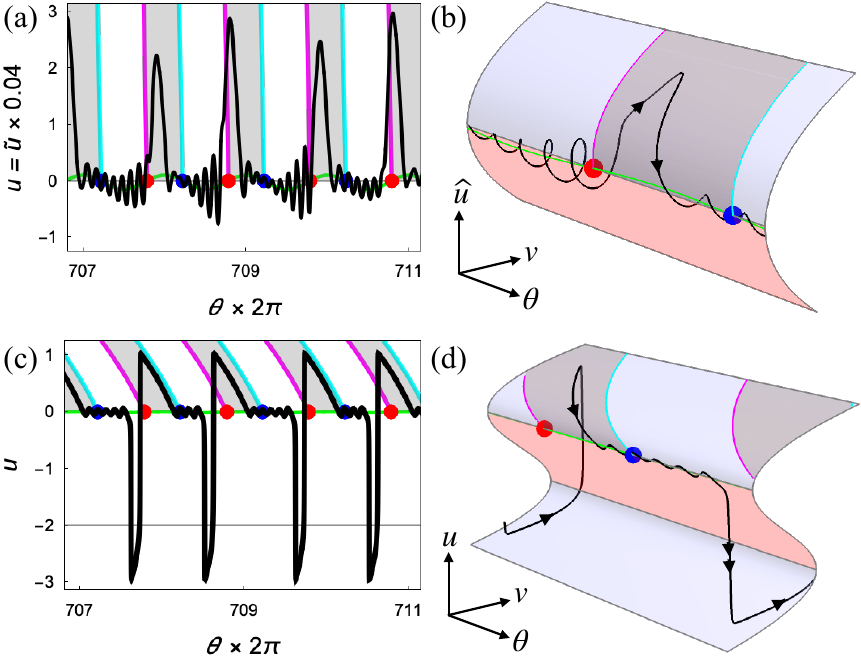}
\caption{
Time traces (black) of the chimera in Fig.~\ref{fig:representatives}, at
$x=0.23$ in (a) and (b), and at $x=0.41$ in (c) and (d). In the decoherent
SAO regions, the time traces oscillate about the weak canards of the FNs
(blue dots) and the faux canards of the FSs (red dots), near the
attracting (upper) sheet and repelling (middle) sheet of the slow
manifold. In the coherent MMO regions, the time traces also have LAOs down
to the lower branch of the attracting slow manifold.
}
\label{fig:xtrace-slowmanifolds}
\end{figure}

The FNs and FSs
lie on a fold curve
between attracting 
and repelling sheets
of the slow manifold.
For \eqref{eq:fvdPol-PDE},
they are located
at the points $\theta_{N,S}$
where $a+b \cos ( \theta_{N,S}) = 0$,
for all $0 < \vert a \vert < b$,
as may be calculated from the ODE consisting 
of the reaction terms in \eqref{eq:fvdPol-PDE},
see Refs.~\onlinecite{BDGKKV2016,FNFS}.
See Fig.~\ref{fig:xtrace-slowmanifolds}.

In the decoherent regions,
the time traces of the SAOs are observed to pass through
neighbourhoods of FNs and FSs in alternation,
Fig.~\ref{fig:xtrace-slowmanifolds}(a,b).
Near a FN,
the SAOs
are centred on its weak canard \cite{SW2004,W2005}.
As the time traces approach the next FS,
the SAOs are centred about
its faux canard \cite{WMR2013}.
The weak and faux canards 
(approximated by the green curves)
lie ${\cal O}(b)$-close to the fold curve,
as expected from the ODE
\cite{VW2015}.
In the passage from an FN to an FS
and then through a neighbourhood of the FS, 
the amplitude grows 
until there is an excursion
(near the repelling branch of the slow manifold),
into the funnel region (gray) of the next FN. 
Hence, the time trace enters the neighbourhood of the next FN,
since the funnels are basins of attraction for the FNs \cite{RW2007},
and there it has further SAOs
about the weak canard of that FN.

In the coherent region(s), the time traces are seen to be MMOs induced by
the canards of the folded singularities
\cite{RW2007,TTVWB2011,HKWS2011,DGKKOW2012}
(Fig.~\ref{fig:xtrace-slowmanifolds}(c,d)). The SAO segments of these MMOs
are also about the FN weak canard. However, here, after passage near a FN,
there is a large-amplitude excursion toward the lower attracting sheet of
the slow manifold. Subsequently, the traces crawl along this sheet until
they reach the lower fold curve, where they jump back to the upper sheet
and into the funnel of the next FN, and the cycle repeats. (The funnel is
enclosed by the true canard (magenta) of the FS
\cite{WMR2013,WMR2013,BBU2020}, the strong canard (cyan) of the FN
\cite{W2005,RW2007}, and the fold curve.) These MMOs remain spatially
coherent in the presence of diffusion, due to the LAOs.

\begin{figure}[h!]
\centering
\includegraphics[width=\columnwidth]{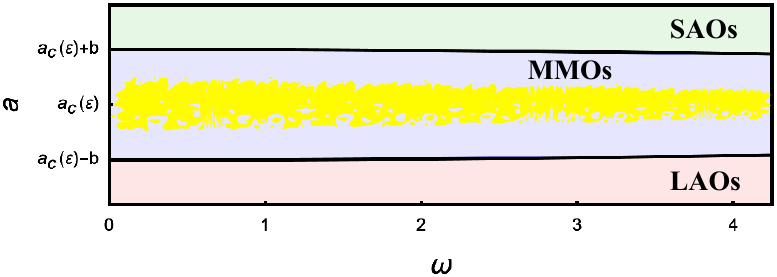}
\caption{
Mixed-amplitude chimeras of \eqref{eq:fvdPol-PDE} are robust in the
goldilocks zone (gold) of the $(\omega,a)$ plane. It sits in the middle of
the MMO region, which is enclosed by $a = a_c(\eps) \pm b e^{- \varepsilon
\omega^2/2 }$, {\it i.e.,} by the saddle-node bifurcations of the FN and
FS canards \cite{BDGKKV2016}. 
For the classical van der Pol equation, $a_c(\eps) =
-\frac{\eps}{8}-\frac{3\eps^2}{32} + \mathcal{O}(\eps^3)$ \cite{BCDD1980}.
}
\label{fig:ODE_aomega}
\end{figure}

Overall, we find that
there is a robust zone
--which we label as a goldilocks zone--
in which the mixed-amplitude chimera states exist in \eqref{eq:fvdPol-PDE}.
See Fig.~\ref{fig:ODE_aomega}. 
In this goldilocks zone,
the distances between the FNs and FSs are simultaneously small
and large enough.
The distances are small
enough that, on wide intervals, the attractor of
\eqref{eq:fvdPol-PDE} can pass near the FNs and FSs in alternation, with
the amplitude remaining small and diffusion causing the decoherence. They
are also large enough that, in complementary MMO regions, the return
mechanisms for the LAO segments are robust.

The goldilocks zone 
shown in Fig.~\ref{fig:ODE_aomega}
is representative 
of the zones 
in the $(\omega,a)$ plane
for a range of values of $b$.
Hence, mixed-amplitude chimeras
are robust in \eqref{eq:fvdPol-PDE}.

\section{Delayed Hopf bifurcations underlie the decoherent SAOs  \label{sec:DHB}}


Having presented the novel mixed-amplitude chimeras,
the parameter regime in which they exist,
the bifurcations in which they are created or annihilated,
and the folded nodes and folded saddles which 
make it possible for these identical oscillators
to exhibit both SAO and LAO/MMO dynamics,
we next turn to a fundamental question about 
the SAO dynamics.
Namely, what is the mechanism
by which the solutions in the decoherent regions
can remain close to the fold curve long enough so that
they can pass through the neighbourhoods
of both the FN and the FS, hence keeping the oscillation
amplitudes small?
This question is important, since it is expected that
solutions which pass through a
neighbourhood of a FN generally
are repelled away from the fold curve
and rapidly approach the (hyperpolarized) attracting branch of 
the slow manifold, resulting in large-amplitudes.
It turns out, as we show here,
that the SAOs in the decoherent regions can exist
due to delayed passage through Hopf bifurcations (DHB).
That is, DHB makes it possible for large measures
of solutions to stay near the fold curve (and hence maintain small amplitude)
so that they not only pass through a neighbourhood of the FN,
but also stay close enough
to pass through a neighbourhood of the FS and into the funnel of the next FN,
so that the cycle can repeat.

To show this, we study \eqref{eq:fvdPol-PDE}
with parameters chosen so that the FN and the FS are
close to $\theta=0 \ {\rm mod} \ 2\pi$.
This includes the parameters in the goldilocks zone.

First, we show that the local dynamics of \eqref{eq:fvdPol-PDE}
may be studied using two coupled CGL-type equations
with a slowly-varying parameter.
In particular, we rescale
$u = \sqrt{\eps} u_2$,
$v = {\eps} v_2$,
$\theta  = {\eps}^\frac{1}{4} \theta_2$,
$\tau=\sqrt{\eps}t$,
$a+b= \sqrt{\eps} \alpha_2$, and
$D_i = \sqrt{\eps} d_i$ ($i=u,v$),
so \eqref{eq:fvdPol-PDE} becomes
\begin{equation}
\begin{split}
{u_2}_{\tau} &= {v_2} - u_2^2 - \tfrac{1}{3} \sqrt{\eps}u_2^3 + d_u {u_2}_{xx} \\
{v_2}_{\tau} &= - u_2 +\alpha_2  -\tfrac{1}{2} b \theta_2^2
+ {\cal O}(\sqrt{\eps}\theta_2^4)
+ d_v {v_2}_{xx} \\
{\theta_2}_{\tau} &= \eps^\frac{1}{4} \omega.
\end{split}
\nonumber
\end{equation}
This rescaling focuses on the dynamics
in the neighborhood of a FN-FS pair.
This local model consists of a fast-slow system,
with $(u_2,v_2)$ as the fast variables
and $\theta_2$ as the slow variable.

We rectify the
critical manifold 
to the $\theta_2$-axis
using ${\widetilde u_2} = u_2 + \mu$
and ${\widetilde v_2} = v_2 - \mu^2$,
where $\mu=-\alpha_2 + \frac{b}{2}\theta_2^2$.
Also, we switch
to the complex coordinates 
$z= {\widetilde u_2} + i {\widetilde v}_2$
and ${\bar z} = {\widetilde u_2} - i {\widetilde v}_2$,
which are more natural variables 
for studying the slow passage through a Hopf bifurcation.
Hence, \eqref{eq:fvdPol-PDE}
is locally equivalent to
\begin{equation*}
\begin{split}
\begin{bmatrix} z \\ \overline{z} \end{bmatrix}_\tau
&= \begin{bmatrix} \mu-i & \mu \\ \mu & \mu+i \end{bmatrix}
\begin{bmatrix} z \\ \overline{z} \end{bmatrix}
- Q \begin{bmatrix} 1 \\1 \end{bmatrix}
+ \begin{bmatrix}  d_+ & d_- \\ d_- & d_+\end{bmatrix}
\begin{bmatrix} z \\ \overline{z} \end{bmatrix}_{xx} \\
{\mu}_{\tau} &= \eps^\frac{1}{4} \omega \sqrt{2b (\mu+\alpha_2)},
\end{split}
\end{equation*}
where $Q= \tfrac{1}{4} (z + {\bar z})^2$ 
and $d_{\pm} = \tfrac{1}{2}(d_u \pm d_v)$.
The origin is an attracting spiral 
of the $(z,{\bar z})$ system
for each $\mu <0$
and a repellor for each $\mu >0$.

Due to the slow growth in $\mu$,
this PDE exhibits a dynamic Hopf bifurcation at $\mu=0$.
In particular,
solutions initialised with $\mu<0$
do not immediately diverge from the equilibrium 
when $\mu$ crosses zero and becomes positive.
Instead, they remain near the repelling state for long times
beyond the dynamic Hopf point (data not shown),
in a delayed Hopf bifurcation (DHB).
The time of the delay, past $\mu=0$,
is $\mathcal{O}(1)$ in the slow time,
so that solutions spend long times near the repelling sheet
of the slow manifold in the robust SAO regions.
Moreover, these times are long enough so that 
the solutions can pass near a pair of FN and FS and
reach the basin of attraction (funnel) of the next FN,
keeping the oscillation amplitude small.

\medskip
\noindent
{\bf Remark.}
For this system of coupled CGL-type PDEs, the observed DHB 
is consistent with the recent discovery
of DHB in nonlinear PDEs \cite{KV2018,GKV2020},
including the Hodgkin-Huxley PDE,
Brusselator model, and CGL equation.
See also Ref.~\onlinecite{ADVW2020}.

\medskip
\noindent
{\bf Remark.}
The analysis here of DHB in these PDEs was inspired 
by the ODE analysis in Ref.~\onlinecite{VW2015}.
There, singularities known as
Folded Saddle Nodes of Type 1 (FSN I),
which arise in bifurcations of FNs and FSs, 
are studied in ODEs.
The analysis in Ref.~\onlinecite{VW2015} 
shows that delayed Hopf bifurcation plays a central role
in ODEs near FSN I bifurcations.
Hence, it was natural here to examine the regime
in which the FN and FS of \eqref{eq:fvdPol-PDE}
are close together, 
and to look for DHB also in the PDE \eqref{eq:fvdPol-PDE}.

\section{Bifurcation to coherent-coherent states
\label{sec:coh-coh}}

The above analyses of the folded singularities
(Sec.~\ref{sec:foldedsingularities})
and of the scaling
leading to the coupled CGL-type model for the local dynamics
(Sec.~\ref{sec:DHB})
also apply for larger $b$ (relative to $\eps$), 
outside the main chimera regime.
With stronger forcing (larger $b$),
the attractors also
consist of regions with coherent MMOs
and complementary regions of SAOs.
However, we find that with larger $b$ the pure-SAO states
are coherent in the presence of diffusion,
rather than decoherent as in the mixed-amplitude chimeras.
These coherent-coherent patterns are multi-mode attractors in the sense of Ref.~\onlinecite{VBK2020}.

\begin{figure}[htbp]
\centering
\includegraphics[width=\columnwidth]{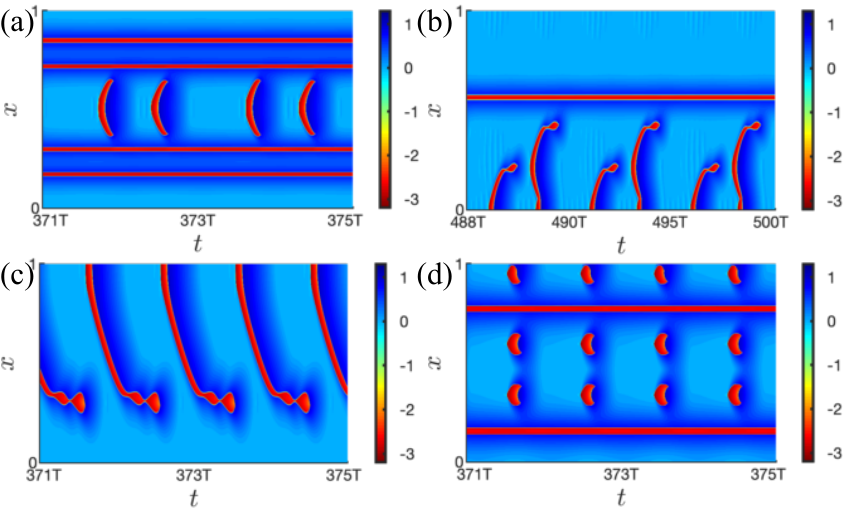} 
\caption{A sample of coherent-coherent multi-mode attractors for various ratios of $\eps$ to $b$.
(a) $\eps = b = 0.005$ with $(D_u,D_v) = (0.47,4)\cdot (\Delta x)^2$.
(b) $\eps = b = 0.01$ with $(D_u,D_v) = (1.71,9) \cdot (\Delta x)^2$.
(c) $\eps = \tfrac{1}{2} b = 0.005$ with $(D_u,D_v)=(1.40625,11.25)\cdot (\Delta x)^2$.
(d) $\eps = \tfrac{1}{5} b = 0.005$ with $(D_u,D_v) = (0.325,8) \cdot (\Delta x)^2$.
Finally, for each fixed pair of $\eps$ and $b$, 
the kinetics parameter $a$ is chosen to lie in the goldilocks zone:
(a) $a = -0.001$, (b) $a=-0.001$, (c) $a= -\tfrac{\eps}{8}+\tfrac{1}{10} b e^{-\eps \omega^2/2}\approx 0.000372503$ and (d) $a= -\tfrac{\eps}{8}+\tfrac{1}{20} b e^{-\eps \omega^2/2} \approx 0.000621879$.
}
\label{fig:CoherentCoherent}
\end{figure}

Examples of these multi-mode attractors, which feature
coherent MMOs and coherent SAOs
are shown in Fig.~\ref{fig:CoherentCoherent}.
Some of these have multiple sharp interfaces separating the regions, 
and others do not have any sharp interfaces. 
For example, in Fig.~\ref{fig:CoherentCoherent}(b),
one finds
an attractor of \eqref{eq:fvdPol-PDE}
with three distinct types of regions:
a region of coherent MMOs,
a region in which the time traces
are near the hyperpolarized state (near $u=-3$),
and a region of coherent SAOs
about the depolarized state.
With the knowledge of the FNs and FSs, 
one can construct rich patterns.

Furthermore,
in these coherent-coherent states
that exist for larger $b$, 
the time traces of the attractor
also exhibit maximal spatiotemporal canards
in the transition intervals
between the two distinct states
(as for example near the bulbs 
in Fig.~\ref{fig:CoherentCoherent}(c)).
These maximal spatiotemporal canards mediate the transitions between regions of distinct oscillatory behaviour, similar to those in Ref.~\onlinecite{VBK2020}.

\section{Conclusions and discussion 
\label{sec:conclusions}}

In this article, 
we presented the novel class of chimeras
labeled as mixed-amplitude chimeras, 
which consist of 
co-existing decoherent SAOs and coherent LAOs or MMOs.  
This constitutes the first observation of chimeras in PDEs
in which the structures, amplitudes, and frequencies 
of the oscillations are widely separated in the two distinct types of regions. 
With the exception of Ref.~\onlinecite{ABE2019},
these chimeras differ from the known chimeras and theory, 
where the amplitudes and frequencies in the coherent and decoherent regions 
are comparable. 

These mixed-amplitude chimeras are surprising, 
since the coherent LAOs and MMOs have much larger amplitudes and
since they are structurally stable.
Moreover, 
in parameter regimes immediately adjacent to the chimera regime,
the LAOs and MMOs are observed 
to invade the decoherent regions 
and wipe out the SAOs.
Indeed, new bifurcations were also discovered in which mixed-amplitude chimeras
emerge from spatially-uniform MMOs, trigger waves, 
and sharp-interface solutions, all of which are ubiquitous 
in bistable PDEs.

Initial theory has been identified for the mixed-amplitude chimeras. 
It was shown here
that they are generated by folded node and folded saddle singularities, 
common in bistable systems. Moreover,
it was discovered that there is a robust
goldilocks zone in which the mixed-amplitude chimeras exist,
where the inter-singularity distances are  simultaneously small and
large enough so the decoherent SAOs and coherent MMOs can co-exist.
Also, it was shown that Delayed Hopf Bifurcations in PDEs 
enables the existence of the SAOs in the decoherent regions.

The mixed-amplitude chimeras may have important applications 
in bistable systems, which are often used to model biological, 
chemical, and physical rhythms. 

The mixed-amplitude chimeras also bear some of the hallmarks of EEG recordings 
taken from animals that exhibit USWS~\cite{Rattenborg2000}. 
Namely, the regions of coherent MMOs consist of large-amplitude, 
low-frequency activity as observed in the sleeping hemisphere, 
and the regions of decoherent SAOs consist of small-amplitude, 
high-frequency activity similar to the awake hemisphere. 

The substantial contrasts in amplitudes and frequencies observed here 
for the mixed-amplitude chimeras of \eqref{eq:fvdPol-PDE} 
are similar in some respects to the contrasts reported recently in
minimal chimeras in Ref.~\onlinecite{ABE2019}.
There, a two-cell model consisting 
of a pair of symmetrically-coupled Lengyel-Epstein ODEs is studied.
The parameters are tuned near a canard point 
of the L-E system,
with one oscillator exhibiting periodic LAOs and the other aperiodic SAOs. 
Moreover, the existence of new types of phase relations between coupled cells 
was established, distinct from the classical in-phase 
and anti-phase attractors in coupled fast-slow systems.
Using analysis similar to that presented in this article, it can be shown that the underlying singularity in Ref.~\onlinecite{ABE2019} is a (type III) folded saddle-node, and the parameters are in its goldilocks zone.

The mixed-amplitude chimeras introduced here
are distinct from the Multi-Mode Attractors
presented in Ref.~\onlinecite{VBK2020}.
There, the MMAs were introduced 
in systems with spatially-dependent parameters.
The spatially-dependent applied currents
(or spatially-dependent maximal ion-channel conductivities)
give rise to attractors that exhibit different modes
of oscillations in different spatial regions.
Hence, in the systems with MMAs,
the oscillators are not all identical.
In contrast, here,
the reactions are all spatially homogeneous
and the diffusion is purely local and isotropic,
so that the oscillators at each point are identical.
See also the discussion in the introductory section 1
of Ref.~\onlinecite{VBK2020}.

\section{Some open questions
\label{sec:openquestions}}

The discovery of mixed-amplitude chimeras raises
a number of questions.
First, it is of interest
to investigate the mechanism(s) by which 
some of the mixed-amplitude chimeras
can have coherent MMOs 
that occur at fixed locations in space
and periodically in time.
These are indicated by the blue circles
in the $D_u - D_v$ parameter plane 
in Fig.~\ref{fig:DuDv-bifurcationdiagram}.
As discussed above, these are distinct 
from the typical chimeras (red circles),
in which the coherent MMOs
occur at random points and at random times.
There may be some type of resonance
responsible for creating these wedges of blue circles
within the sea of red circles.

Second, we ask if some of the known examples of chimeras
in other systems
can be continued to parameter regimes in which the
structures, amplitudes, and/or frequencies of the coherent oscillations 
are substantially different from those of the decoherent SAOs, as
is the case for the mixed-amplitude chimeras here.
The known examples and theory of chimera states
in coupled oscillators,
integro-differential systems, and PDEs
are based on 
the amplitude-phase method or
mean-field approach \cite{KB2002,AS2004,PA2015,O2018,SS2014}, 
symmetry and broken symmetry analyses \cite{KHK2018,ZKS2014}, 
and the theory
\cite{Zhang2021,OMT2008} of desynchronization-induced stabilization of
coherent oscillations,
among other important results.
These methods have been used primarily to find
chimeras in which the amplitudes and frequencies
of the oscillations
in the decoherent and coherent regions are comparable,
{\it i.e.,} where they do not differ by orders of magnitude.
For example, the analysis used 
in the phase-amplitude method 
is premised on the fact that the oscillations
in the chimera states and the states from which bifurcate
have similar amplitudes and frequencies.
Hence, it would of interest to explore 
whether any of these may be continued in parameter space
so that the coherent regions exhibit LAOs and MMOs.

In the other direction,
one may ask wheher the new mixed-amplitude chimeras 
can be continued in parameter space
to a regime in which the coherent
oscillations are near homogeneous states (and hence purely
of SAO type), instead of being MMOs or LAOs.

Third, 
there is a natural question 
about what the minimal dimension is
for a system to exhibit mixed-amplitude chimeras.
We recall that, in ODEs,
three is the minimum number of variables necessary for MMOs.
This is the case for the forced van der Pol reaction
terms in \eqref{eq:fvdPol-PDE}.
It is also the case 
for the Lengyel-Epstein ODEs studied in Ref.~\onlinecite{ABE2019},
where an individual L-E model has two dependent variables,
and the systems studied there consist of
two symmetrically-coupled L-E oscillators ({\it i.e.,} four variables),
or of a single L-E oscillator with some type of external forcing
({\it i.e.,} three variables).
Hence,
the results of Ref.~\onlinecite{ABE2019} and this article suggest that three
variables in the kinetics 
is expected to be the minimum number needed 
for chimera states 
to exhibit co-existing regions of coherent MMOs and
decoherent SAOs,
such as the mixed-amplitude chimeras reported here.


\begin{acknowledgments}
The authors thank Karen Corbett for useful conversations and support with MASSIVE. 
We also thank Irv Epstein, Ryan Goh, Christian Kuehn, Jonathan Touboul, Gene Wayne, and Martin Wechselberger for useful comments and questions.
The research of T.V.
was partially supported by NSF-DMS 1853342.
The research of T.K.
was partially supported by NSF-DMS 1616064.
This work was supported by the MASSIVE HPC facility (www.massive.org.au).
\end{acknowledgments}

\section{Appendix \label{sec:appendix}}

The numerical simulations
were performed
using the method of lines,
with adaptive time-stepping 
suited to stiff problems,
and a second-order spatial discretization.
The results were 
confirmed independently using four other methods:
balanced symmetric Strang operator splitting, 
with (second-order) centered finite differences in space 
and fourth-order Runge-Kutta in time;
the Crank-Nicholson method,
which is second-order accurate in both space and time; 
a compact finite difference method which is fourth-order accurate in space;
and, a Chebyshev grid method that is spectral in space.
All simulations were run 
at least for the order of the diffusive time scale, based on the 
smaller diffusivity.
Unless stated otherwise,
$N_x=750$ spatial subintervals were used
($\Delta x = \frac{1}{750}$).
Also, convergence was checked
with smaller values of $\Delta x$.

The main method used to generate initial data $(u_0(x),v_0(x))$
on $[0,1]$
is to compute a limit cycle attractor $(\Gamma_u(t),\Gamma_v(t))$
(period $T$) of the kinetics ODE,
and then to set $(u_0(x),v_0(x))=
\left( \Gamma_u\left(\frac{t}{nT}\right),
\Gamma_v\left(\frac{t}{nT}\right) \right)$.
Typical initial data consists
of $n$ copies spread out in space,
with $\theta(t=0)=0$.
For example,
$n=6$ in Fig.~\ref{fig:DuDv-bifurcationdiagram}.
Eight other types of initial data were also used 
to explore more fully the multi-stable landscape of solutions,
and to find chimera states.


\begin{thebibliography}{99}

\setlength{\parskip}{0pt}

\small

\bibitem{KB2002}
Y. Kuramoto and D. Battogtokh,
{\it Nonlin. Phen. Complex Sys.} {\bf 5}, 
380 (2002).
Y. Kuramoto,
in {\it Nonlin. Dyn. Chaos:
Where Do We Go from Here?},
S.J. Hogan, A.R. Champneys, B. Krauskopf, M. di Bernardo,
H.M. Osinga, and M.E. Homer (eds.),
IOP, Bristol, UK, 209 (2003).
Y. Kuramoto 
and S.I. Shima,
{\it Prog. Theor. Phys. Suppl.} {\bf 150}, 
115 (2003).
S.I. Shima and Y. Kuramoto,
{\it Phys. Rev. E} {\bf 69}, 
036213 (2004).

\bibitem{AS2004}
D.M. Abrams and S.H. Strogatz,
{\it Phys. Rev. Lett.} {\bf 93}, 
174102 (2004).
D.M. Abrams, R.E. Mirollo, S.H. Strogatz, D.A. Wiley,
{\it Phys. Rev. Lett.} {\bf 101}, 
084103 (2008).

\bibitem{OA2008}
E. Ott and T.M. Antonsen,
{\it Chaos} {\bf 18}, 
037113 (2008).
E. Ott and T.M. Antonsen,
{\it Chaos} {\bf 19}, 
023117 (2009).

\bibitem{OMT2008}
O.E. Omel'chenko, Y.L. Maistrenko, and P.A. Tass,
{\it Phys. Rev. Lett.}
{\bf 100}, 044105 (2008).

\bibitem{BPR2010}
G. Bordyugov, A. Pikovsky,
and M. Rosenblum,
{\it Phys. Rev. E} {\bf 82},
035205(R) (2010).

\bibitem{L2010}
C. Laing,
{\it Phys. Rev. E} {\bf 81},
066221 (2010).

\bibitem{L2015}
C. Laing,
{\it Phys. Rev. E} {\bf 92}, 
050904(R) (2015).
C. Laing,
{\it Phys. Rev. E} {\bf 100}, 
042211 (2019).


\bibitem{OMHS2011}
I. Omelchenko, Y. Maistrenko, P. H\"ovel, and E. Sch\"oll,
{\it Phys. Rev. Lett.} {\bf 106},
234102 (2011).
I. Omelchenko, O. E. Omel'chenko, P. H\"ovel, and E. Sch\"oll,
{\it Phys. Rev. Lett.} {\bf 110},
224101 (2013).

\bibitem{TNS2012}
M.R. Tinsley, S. Nkomo, and K. Showalter,
{\it Nature Phys.} {\bf 8}, 
662 (2012).

\bibitem{HMRHOS2012}
A.M. Hagerstrom, T. E. Murphy,
R. Roy, I. Omelchenko, P. H\"ovel,
and E. Sch\"oll, 
{\it Nature Phys.} {\bf 8}, 
658 (2012).

\bibitem{MTFH2013}
E.A. Martens, S. Thutupalli, A. Fourriere, and O. Hallatschek,
{\it PNAS} {\bf 110}, 
10563 (2013).

\bibitem{NTS2013}
S. Nkomo, M.R. Tinsley, and K. Showalter,
{\it Phys. Rev. Lett.} {\bf 110}, 
244102 (2013).

\bibitem{SZHK2014}
K. Sch\"onleber, C. Zensen, A. Heinrich, and K. Krischer,
{\it New J. Phys.} {\bf 16},
063024 (2014).
L. Schmidt, K. Sch\"onleber, K. Krischer, and V. Garcia-Morales,
{\it Chaos} {\bf 24}, 
013102 (2014).
L. Schmidt and K. Krischer,
{\it Chaos} {\bf 25},
064401 (2015).

\bibitem{SS2014}
G.C. Sethia and A. Sen,
{\it Phys. Rev. Lett.}
{\bf 112}, 144101 (2014).



\bibitem{YPR2014}
A. Yeldesbay, A. Pikovsky, and M. Rosenblum,
{\it Phys. Rev. Lett.} {\bf 112},
144103 (2014).

\bibitem{ZKS2014}
A. Zakharova, M. Kapeller, and E. Sch\"oll,
{\it Phys. Rev. Lett.} {\bf 112},
154101 (2014).

\bibitem{OPHSH2015}
I. Omelchenko, A. Provata, J. Hizanidis, E. Sch\"oll, and P. H\"ovel,
{\it Phys. Rev. E} {\bf 91},
022917b (2015).

\bibitem{PA2015}
M.J. Panaggio and D.M. Abrams,
{\it Nonlinearity} {\bf 28}, 
R67 (2015).

\bibitem{CCFGR2016}
M.G. Clerc, S. Coulibaly,
M.A. Ferre, M.A. Garcia-Nustes,
and R.G. Rojas,
{\it Phys. Rev. E} {\bf 93},
052204 (2018).

\bibitem{KHSKK2016}
F. P. Kemeth, S. W. Haugland, L. Schmidt, I. G. Kevrekidis, and K. Krischer,
{\it Chaos} {\bf 26},
094815 (2016).

\bibitem{LD2016}
B.-W. Li and H. Dierckx,
{\it Phys. Rev. E} {\bf 93}, 
020202(R) (2016).
B.-W. Li, Y. He, L.-D. Li,
L. Yang, X. Wang,
{\it Arxiv} 2012.00983, nlin, Dec 2020.




\bibitem{NRM2017}
Z.G. Nicolau, H. Riecke, and A.E. Motter, 
{\it Phys. Rev. Lett.} {\bf 119}, 
244101 (2017).

\bibitem{KHK2018}
F.P. Kemeth, S.W. Haugland,
and K. Krischer,
{\it Phys. Rev. Lett.}
{\bf 120}, 214101 (2018).

\bibitem{O2018}
O.E. Omel'chenko,
{\it Nonlinearity} {\bf 31}, 
R121 (2018).


\bibitem{TRTSE2018}
J.F. Totz, J. Rode, M.R. Tinsley,
K. Showalter, and H. Engel,
{\it Nat. Phys.} {\bf 14}, 
282 (2018).
J. Rode, J.F. Totz, E. Fengler, and H. Engel,
{\it Frontiers Appl. Math. Stat.} {\bf 5}, 
31 (2019).
J.F. Totz, M.R. Tinsley,
H. Engel, and K. Showalter,
{\it Nat. Sci. Rep.} {\bf 10}, 
7821 (2020).

\bibitem{KMMG2018}
S. Kundu, S. Majhi,
P. Muruganandam,
D. Ghosh,
{\it Eur. Phys. J. Spec. Topics} {\bf 227},
983--993 (2018).

\bibitem{ABE2019}
N.M. Awal, D. Bullara, and I.R. Epstein,
{\it Chaos} {\bf 29}, 
013131 (2019).
N.M. Awal and I.R. Epstein,
{\it Phys. Rev. E} {\bf 101}, 
042222 (2020).



\bibitem{SA2021}
I.A. Shepelev and V.S. Anischenko,
{\it Comm. Nonlin. Sci., Num. Sim.}
{\bf 93}, 105513 (2021).

\bibitem{Zhang2021}
Y. Zhang and A.E. Motter,
{\it Phys. Rev. Lett.} {\bf 126}, 
094101 (2021).


\bibitem{BKR2008}
M. Brons, T.J. Kaper, and H.G. Rotstein,
{\it Chaos} {\bf 18}, 015101
(2008).

\bibitem{DGKKOW2012}
M. Desroches, J. Guckenheimer, B. Krauskopf, 
C. Kuehn, H.M. Osinga, and M. Wechselberger
{\it SIAM Rev.} {\bf 54}, 211
(2012).

\bibitem{HKWS2011}
E. Harvey, V. Kirk, M. Wechselberger, and J. Sneyd,
{\it J. Nonlin. Sci.}, {\bf 21} 639--683 (2011).

\bibitem{RW2007}
J.E. Rubin and M. Wechselberger,
{\it Bio. Cybernetics} {\bf 97}, 
5 (2007). 
M. Br{\o}ns, M. Krupa, and M. Wechselberger,
{\it Fields Inst. Commun.} {\bf 49}, 
39-63 (2006).



\bibitem{K1995}
M. Koper,
{\it Physica D} {\bf 80}, 
72 (1995).




\bibitem{W1997}
D. Walgraef,
{\it Spatio-Temporal Pattern Formation},
Springer, New York (1997).

\bibitem{vS2003}
W. van Saarloos,
{\it Phys. Rep.},
{\bf 386}, 29 (2003).

\bibitem{N2002}
Y. Nishiura,
{\it Far-from Equilibrium Dynamics,}
Amer. Math. Soc.,
Transl. Math. Monographs, {\bf 209}
(2002).

\bibitem{KS2009}
J. Keener and J. Sneyd,
{\it Mathematical Physiology I: Cellular Physiology},
2nd ed., Springer, New York (2009).

\bibitem{HH1952}
A. Hodgkin and A. Huxley,
{\it J. Physiol.} {\bf 117}, 
500 (1952).

\bibitem{M2006}
J. Moehlis,
{\it J. Math. Bio.} {\bf 52}, 
141 (2006).

\bibitem{F1960}
R. FitzHugh,
{\it J. Gen. Physiol.} {\bf 43},  
867 (1960).

\bibitem{NAY1962}
J. Nagumo, S. Arimoto, and S. Yoshizawa,
{\it Proc. IRE} {\bf 50}, 
2061 (1962).

\bibitem{AP1996}
R.R. Aliev and A.V. Panfilov,
{\it Chaos Soliton Fract}
{\bf 7}, 293 (1996).

\bibitem{AFS2004}
D. Angeli, J.E. Ferrell, and E.D. Sontag,
{\it PNAS}, 
{\bf 101}, 1822 (2004).

\bibitem{RG2021}
J. Rombouts and L. Gelens,
{\it PLOS Comp. Bio.},
{\bf 17}, 1008231 (2021).

\bibitem{LRE1990}
I. Lengyel, G. Rabai, and I.R. Epstein,
{\it J. Am. Chem. Soc.} {\bf 112}, 
9104 (1990).
I. Lengyel and I.R. Epstein,
{\it Science} {\bf 251}, 
650 (1991).



\bibitem{EM2008}
I. Erchova and D.J. McGonigle,
{\it Chaos} {\bf 18}, 
015115 (2008).

\bibitem{DBSK2007}
C.G. Diniz Behn, E.N. Brown, T.E. Scammell, and N. Kopell,
{\it J. Neurophysiol.} {\bf 97}, 
3828-3840 (2007).

\bibitem{Wang2010}
X.-J. Wang,
{\it Physiol. Rev.} {\bf 90}, 
1195-1268 (2010).

\bibitem{Motter2010}
A.E. Motter,
{\it Nat. Phys.}
{\bf 6}, 164--165 (2010).

\bibitem{Ramlow2019}
L. Ramlow, J. Sawicki, A. Zakharova, J. Hlinka, J.C. Claussen, and E. Sch\"{o}ll,
{\it EPL}
{\bf 126}, 50007 (2019).

\bibitem{Haugland2015}
S.W. Haugland, L. Schmidt, and K. Krischer,
{\it Sci. Rep.-UK}
{\bf 91}, 40006 (2010).

\bibitem{Ma2010}
R. Ma, J. Wang, and Z. Liu,
{\it EPL}
{\bf 91}, 40006 (2010).

\bibitem{Rattenborg2000}
N.C. Rattenborg, C.J.Amlaner, and S.L. Lima,
{\it Neurosci. Behav. Rev.}
{\bf 24}, 817--842 (2000).

\bibitem{Rattenborg2006}
N.C. Rattenborg,
{\it Naturwissensch.}
{\bf 93}, 413--425 (2006).

\bibitem{SMS1993}
M. Steriade, D.A. McCormick, and T.J. Sejnowski,
{\it Science} {\bf 262}, 
679-685 (1993).


\bibitem{vdP1920}
B. van der Pol,
{\it Radio Review} {\bf 1}, 701 (1920).
B. van der Pol,
{\it Lond., Edinb., Dublin Phil. Mag. \& J. Sci., Series 7},
{\bf 3},
65--80 (1927).

\bibitem{VKKR1987}
A.B. Vasil'eva, S.A. Kashchenko, Yu.S. Kolesov, and N.Kh. Rozov,
{\it Math. USSR Sbor.} {\bf 58}, 491
(1987).

\bibitem{SE1988}
K.R. Schneider and V.M. Evtukhov,
{\it Differentsial'nye  Urav.} {\bf 24}, 
1027 (1988)
(Eng. transl.: {\it Diff. Eq.} {\bf 24}, 683)

\bibitem{KG1995}
D. Kaplan and L. Glass,
{\it Understanding Nonlinear Dynamics,}
Springer, New York (1995).

\bibitem{FLB2002}
M.L. Fachinetti, E. de Langre, and F. Biolley
{\it Comp. Rend. Acad. Sci., Paris, Ser. II} {\bf 330} 
(2002).

\bibitem{GHW2003}
J. Guckenheimer, K. Hoffman, and W. Weckesser,
{\it SIAM J. Appl. Dyn. Sys.} {\bf 2},
1 (2003);
K. Bold, C. Edwards, J. Guckenheimer, S. Guharay,
K. Hoffman, J. Hubbard, R. Oliva, and W. Weckesser,
{\it ibid.} {\bf 2},
570 (2003).

\bibitem{VW2015}
T. Vo and M. Wechselberger,
{\it SIAM J. Math. An.} {\bf 47}, 
3235 (2015).

\bibitem{BDGKKV2016}
J. Burke, M. Desroches, A. Granados,
T.J. Kaper, M. Krupa, and T. Vo,
{\it J. Nonlin. Sci.} {\bf 26}, 
405 (2016).

\bibitem{SW2004}
P. Szmolyan and M. Wechselberger,
{\it J. Diff. Eq.} {\bf 200},
69 (2004).

\bibitem{HB2012}
X. Han and Q. Bi,
{\it Nonlinear Dynamics} {\bf 68},
275 (2012).

\bibitem{form-fvdPol}
The kinetics
in \eqref{eq:fvdPol-PDE}
are equivalent
to the classical
$x_t=y - \left( \frac{1}{3} x^3 - x \right)$,
$y_t=\eps(\tilde{a}-x +b \cos(\theta))$
under the linear shift
$u=x-1$,
$v=y+\frac{2}{3}$,
and $a = \tilde{a}-1$.



\bibitem{BDE1995}
F. Buchholtz, M. Dolnik, and I.R. Epstein,
{\it J. Phys. Chem.} {\bf 99}, 
15093 (1995).

\bibitem{TTVWB2011}
W. Teka, J. Tabak, T. Vo, M. Wechselberger, and R. Bertram,
{\it J. Math. Neuro.} {\bf 1},
12 (2011).

\bibitem{WMR2013}
M. Wechselberger, J. Mitry, and J. Rinzel,
in {\it Nonautonomous Dynamical Systems in the Life Sciences},
{\bf LNM 2012}, Springer,
89 (2013).
J. Mitry and M. Wechselberger,
{\it SIAM J. Appl. Dyn. Sys.} {\bf 16}, 
546 (2017).

\bibitem{KAA2014}
M. Krupa, B. Ambrosio, and M.A. Aziz-Alaouio,
{\it Nonlinearity} {\bf 27}, 
1555 (2014).

\bibitem{RRW2015}
K.-L. Roberts, J.E. Rubin, and M. Wechselberger,
{\it SIAM J. Appl. Dyn. Sys.} {\bf 14}, 
1808 (2015).


\bibitem{BBU2020}
E. Bossolini, M. Br{\o}ns, and K. Uldall-Kristiansen,
{\it SIAM Rev.} {\bf 62}, 869 (2020).


\bibitem{BBE1991}
M. Br{\o}ns and K. Bar-Eli,
{\it J. Phys. Chem} {\bf 95},
8706 (1991).

\bibitem{MKP2014}
P. de Maesschalck, E. Kutafina, and N. Popovic,
{\it J. Dyn. Diff. Eq.} {\bf 26}, 
955 (2014).

\bibitem{W2005}
M. Wechselberger,
{\it SIAM J. Appl. Dyn. Sys.} {\bf 4}, 
101 (2005).
M. Krupa and M. Wechselberger,
{\it J. Diff. Eq.} {\bf 248}, 
2841 (2010).



\bibitem{FNFS}
The local minimum
of the critical manifold
$\{ (u,v,\theta) \vert v=u^2 + \frac{1}{3} u^3\}$,
for $\eps=0$.
On this manifold,
the reduced flow is 
$u(u+2) \dot{u} = a - u + b\cos(\theta)$ and
$\dot{\theta} = \omega$,
as derived by taking $\frac{d}{dt}$
of the algebraic equation defining the manifold.
The regular FN dynamics occur for 
$b^2 < a^2 + \tfrac{1}{64 \omega^2}$.

\bibitem{BCDD1980}
E. Benoit, J. L. Callot, F. Diener, and M. Diener, 
{\it Collectanea Mathematica} {\bf 31}, 
3 (1980).
W. Eckhaus, in {\it Lecture Notes in Math.} {\bf 985}, 
Springer-Verlag, 449 (1983).
A.K. Zvonkin and M.A. Shubin,
{\it Russ. Math. Surv.} {39}, 
69 (1984).
M. Br{\o}ns,
in {\it Proc. 9th AIMS Intl. Conf.,}
{\it Disc. Cont. Dyn. Sys., Suppl.} 77-83 (2013).

\bibitem{KV2018}
T.J. Kaper and T. Vo,
{\it Chaos} {\bf 28}, 
091103 (2018).

\bibitem{GKV2020}
R. Goh, T.J. Kaper, and T. Vo,
ArXiv DS:2012.10048
(2020).

\bibitem{ADVW2020}
D. Avitabile, M. Desroches, R. Veltz, and M. Wechselberger,
{\it SIAM J. Math. An.}
{\bf 52}, 5703 (2020).

\bibitem{VBK2020}
T. Vo, R. Bertram, and T. J. Kaper,
{\it Physica D} {\bf 411}, 132544 (2020).




\end{thebibliography}

\end{document}